\documentclass[twocolumn,preprintnumbers,amsmath,amssymb]{revtex4}
\usepackage{graphicx}% Include figure files
\usepackage{dcolumn}% Align table columns on decimal point
\usepackage{bm}% bold math
\begin{document}
\title{Lorentz, Edwards transformations and the principle of permutation invariance}

\author{Jian Qi Shen}

\email{jqshen@coer.zju.edu.cn}

\affiliation{Zhejiang Institute of Modern Physics and Department
of Physics, Zhejiang University (Yuquan Campus), Hangzhou 310027,
People's Republic of China}

\date{\today}

\begin{abstract}
The Lorentz transformation is derived without the postulate of the
universal limiting speed, and the general Edwards transformation
is obtained by using the principle of permutation invariance
(covariance). It is shown that the existences of the one-way
universal limiting speed (in the Lorentz transformation) and the
constancy of the two-way average speed of light (in the Edwards
transformation) are the necessary consequences of the principle of
permutation invariance that is consistent with the postulate of
relativity. The connection between the Edward transformation and
the general coordinate transformation is discussed, and based on
this, we find that the physical meaning of the Edward parameter,
which indicates the anisotropy of the speed of light, is a
gravitomagnetic potential of the spacetime.
\end{abstract}
%\pacs{}

\maketitle

\section{Introduction}

In the textbook of classical mechanics, field theory and
electrodynamics \cite{Jackson}, the Lorentz transformation is
derived by using Einstein's two postulates. In fact, the Lorentz
transformation can also be obtained without the second postulate,
{\it i.e.}, the existence of a universal limiting speed (constancy
of the speed of light). In other words, the universal limiting
speed can be derived from the purely logical deduction (related to
the principle of relativity). Historically, such a viewpoint was
suggested by some authors \cite{Terletskii,Lee,Mermin,Coleman}. In
this paper, a new way based on the so-called principle of
permutation invariance (covariance) is proposed to confirm our
belief that a priori assumption of a universal limiting speed is
not necessary in the derivation of spacetime coordinate
transformations ({\it e.g.}, Lorentz and Edwards transformations
\cite{Edwards}), and on the contrary, the existence of the
universal limiting speed is a necessary consequence of the
postulate of relativity. Furthermore, by using the principle of
permutation invariance, we generalize the Lorentz and Edwards
transformation to a more general form that agrees with the
postulate of relativity. We show how far such a generalization can
go under the condition that the generalization should fulfill the
principle of relativity.

It is worth emphasizing that the one-way speed of light is not an
observable quantity, since when we measure the one-way speed
between two distant points, the time measurement involves two
clocks, which should have been synchronized first and then placed
at these two points \cite{Edwards}. However, this needs a priori
assumption of a universal limiting speed ({\it e.g.}, a finite,
constant speed or an infinitely large velocity \cite{Finberg}).
Someone may suggest that we can make use of the means of
synchronizing clocks by slow transport, {\it i.e.}, we first set
two clocks at one point (say, A), and then carry one clock very
slowly to another point (say, B) \cite{Ellis}. But such a
synchronization might be affected by carrying the clock from point
A to point B. If the effect of the motion upon the translated
clock was known, then this method might be feasible
\cite{Edwards}. This means that the postulate of constancy of the
one-way speed of light lack the physical foundations
\cite{Edwards}. However, the two-way average speed of a light from
point A travelling to point B and by reflection, back to A, is an
observable quantity since here only one clock is involved, and the
clock synchronization is not required. In other words, the
experiments (such as Michelson-Morley experiment), which were said
to verify the postulate of constancy of the speed of light,
actually demonstrates that only the two-way speed of a light (in a
closed path of given length) is invariant, and the constancy of
the one-way speed of light has so far never been examined in
experiments  \cite{Edwards}. The Edwards transformation, which was
suggested in 1963 based on the postulate of the constancy of the
average speed of light, is a generalization of the Lorentz
transformation. In this transformation, the speed of light in free
vacuum is anisotropic, but the average speed of light in a two-way
(round-trip) process is a constant number (independent of
reference frames). Although there has so far never been any
experimental evidences for the anisotropy of the speed of light,
from the point of view of physical logic, the Edwards
transformation is a self-consistent coordinate transformation.

In the above, we pointed out that there exist no absolute
definitions of simultaneity. In what follows, we discuss in detail
the possibility of the arbitrariness of simultaneity within the
framework of special relativity. The relativity of simultaneity
may be the most important quintessence in special relativity. In
the course of teaching special relativity to undergraduates, the
special relativity is conveyed often by using Einstein's
definition of simultaneity (usually called standard synchrony).
But this is not the unique existing example of relativity of
simultaneity, because the principle of causality allows the
existence of other alternative definitions of simultaneity
\cite{Reichenbach,Grunbaum}. In an Einsteinian universe, no causal
influence can travel faster than the light in free vacuum.
However, Reichenbach suggested that there is no reason to rule out
the possibility of arbitrarily fast causal influences, which would
then be able to single out a unique event at point A that would be
simultaneous with the event at point B, because the unique
standard to determine the time sequence is such that all the
consequences follow the causations \cite{Reichenbach}. If a pulse
travels from point A at time $t_{1}$ to point B and by flection,
back to point A at time $t_{3}$, then the time $t_{2}$ when the
pulse arrives at point B (measured by the clock at point B) can be
said to be simultaneous with the time
$t_{1}+\epsilon(t_{3}-t_{1})$ (measured by the clock at point A),
where $\epsilon \in [0, 1]$. It can be verified that as long as
$\epsilon \in [0, 1]$, such a $t_{2}$ will not violate the
principle of causality \cite{Reichenbach}. In particular,
Einstein's definition of simultaneity (standard synchrony)
corresponds only to a special case, where $\epsilon=1/2$. Here
Einstein's standard synchrony is equivalent to the requirement
that the one-way speeds of the light be the same on the two
segments of its round-trip journey between points A and B. Someone
may argue that Einstein's definition is the only possible choice
to define the relativity of simultaneity. In fact, within the
framework of special relativistic physics, other alternative
choices ($\epsilon\neq 1/2$), although perhaps less convenient,
are indeed possible to give the self-consistent definitions of
simultaneity. The value of $\epsilon$ depends upon the means of
clock synchronization. If we have a pulse signal with infinite
velocity, then we can determine the value for $\epsilon$. But in
fact there might exist no such pulse signals, and it is therefore
impossible within the framework of special relativity for any
synchrony methods to result in fixing any particular value of
$\epsilon$ to the exclusion of any other particular values
\cite{Winnie}. In other words, the one-way speed of light cannot
be determined and is not an observable quantity. This means that
the definition of simultaneity is in a sense arbitrary, and that
Einstein's simultaneity is only the most simplest one among all
the definitions of simultaneity based on various means of clock
synchronization \cite{Zhang}. Different means of clock
synchronization can yield different effects on the quantities such
as the one-way velocity and the simultaneity, which are not
directly observable. But they will not affect all the observable
quantities such as the two-way average speed of light.

As there are various possible means of clock synchronization, and
some alternative definitions of simultaneity different from
Einstein's standard synchrony will not violate the principle of
causality, we can suggest a concept of {\it synchronization gauge}
to indicate the arbitrariness in the definition of relativity of
simultaneity. We show that the essence of the synchronization
gauge is just the general coordinate transformation. In the
present paper, we use the so-called principle of permutation
invariance (covariance), which is consistent with the postulate of
relativity, to derive the Lorentz and Edwards transformations. The
postulate of the universal limiting speed is no longer required in
the derivation of the Lorentz transformation. Instead, it is a
theoretical consequence of the principle of permutation
invariance. In the derivation of the Edwards transformation, the
constancy of the two-way average speed of light in a closed path
can be derived by using the permutation operation. The connection
between the Edwards transformation and the general coordinate
transformation is considered with special emphasis on the physical
meaning of the Edwards parameter that indicates the anisotropy of
the one-way speed of light. It will be shown that the Edwards
spacetime is different from the Minkowski spacetime only by a
global coordinate transformation.

\section{Derivation of the Lorentz transformation}

\subsection{Coordinate transformation without light}
In the derivation of the Lorentz transformation, we first consider
the motion of equation of an ordinary test particle rather than of
the photon, and then set zero for the derivative of the
transformation coefficient with respect to the particle velocity
(the transformation coefficient should be independent of the
motion of the test particle). Thus a preliminary form of the
coordinate transformation is achieved.

As in standard derivation of the Lorentz transformation, we
consider two inertial reference frames (K and K$'$) with spacetime
coordinates $(x, y, z, t)$ and $(x', y', z', t')$, respectively,
and moving relative to each other with a relative velocity $v$
along $\hat{x}$-direction, the most simple linear coordinate
transformation may be as follows
\begin{equation}
\left\{
\begin{array}{ll}
&  x'=k(x-vt),                                 \\
&   x=k(x'+vt').
\end{array}
\right. \label{linear3}
\end{equation}
Here, for convenience, we consider the transformation of the 1+1
dimensional spacetime only. We assume that a test particle is
moving in frame K with velocity $u$ and in primed frame K$'$ with
velocity $u'$. The equations of motion of this particle in frames
K and K$'$ are therefore $x=ut$ and $x'=u't'$, respectively,
provided that the initial location at $t=t'=0$ in frames K and
K$'$ are coincident at the origins of K and K$'$. Substitution of
these two equations into Eq. (\ref{linear3}) yields
\begin{equation}
\frac{1}{k^{2}}=\left(1-\frac{v}{u}\right)\left(1+\frac{v}{u'}\right).
\label{k2}
\end{equation}
Since the transformation coefficient $k$ is independent of the
velocities $u$ and $u'$ of the test particle, one can have
\begin{equation}
\frac{{\rm d}\frac{1}{k^{2}}}{{\rm d}u}=0.    \label{derivative}
\end{equation}
From Eq. (\ref{derivative}), one can obtain the following equation
\begin{equation}
\frac{{\rm d}u'}{{\rm d}u}=\frac{u'^{2}+u'v}{u^{2}-uv},
\label{differential}
\end{equation}
the solution to which is of the form
\begin{equation}
u'=\frac{u-v}{1+\lambda u},      \label{lambda}
\end{equation}
where $\lambda$ is a parameter, which is independent of the
variable $u$. Inserting expression (\ref{lambda}) into Eq.
(\ref{k2}), one can obtain
\begin{equation}
k^{2}=\frac{1}{1+\lambda v}.     \label{ksquared}
\end{equation}
It follows from Eqs. (\ref{linear3}) and (\ref{ksquared}) that the
preliminary form of the spacetime coordinate transformation is
\begin{equation}
\left\{
\begin{array}{ll}
&  x'=\frac{1}{\sqrt{1+\lambda v}}(x-vt),                                 \\
&   t'=\frac{1}{\sqrt{1+\lambda v}}(t+\lambda x),
\end{array}
\right. \label{linear4}
\end{equation}
and the inverse transformation is
\begin{equation}
\left\{
\begin{array}{ll}
&  x=\frac{1}{\sqrt{1+\lambda v}}(x'+vt'),                                 \\
&   t=\frac{1}{\sqrt{1+\lambda v}}(t'-\lambda x').
\end{array}
\right. \label{linear5}
\end{equation}
It should be noted that the inverse transformation can also be
obtained by the permutation operation, which will be discussed in
detail when deriving the Edwards transformation. Apparently,
transformations (\ref{linear4}) and (\ref{linear5}) are not
explicit forms, where the $\lambda$ parameter should be
determined.

\subsection{Coordinate transformations among three inertial frames of reference}
In an attempt to determine the $\lambda$ parameter involved in
Eqs. (\ref{linear4}) and (\ref{linear5}), we introduce a third
reference frame, and then discuss the completeness condition of
the transformations. The third reference frame is K$''$ that is
moving relative to frame K$'$ with a relative velocity $\omega$ in
the positive $\hat{x}$-direction. Thus there are three coordinate
transformations among these three inertial reference frames K,
K$'$ and K$''$. In order to distinguish the $\lambda$ parameters
involved in the transformations, Eq. (\ref{linear4}) (the
spacetime coordinate transformation from K to K$'$) can be
rewritten as
\begin{equation}
\left\{
\begin{array}{ll}
&  x'=\frac{1}{\sqrt{1+\lambda_{1} v}}\left(x-vt\right),                                 \\
&   t'=\frac{1}{\sqrt{1+\lambda_{1} v}}\left(t+\lambda_{1}
x\right).
\end{array}
\right. \label{linear7}
\end{equation}
Assume that the space and time coordinate of a point in frame
K$''$ is $(x'', t'')$. Then the second transformation (from K$'$
to K$''$) is given by
\begin{equation}
\left\{
\begin{array}{ll}
&  x''=\frac{1}{\sqrt{1+\lambda_{2} \omega}}(x'-\omega t'),                                 \\
&   t''=\frac{1}{\sqrt{1+\lambda_{2} \omega}}(t'+\lambda_{2} x').
\end{array}
\right. \label{linear8}
\end{equation}
Substituting Eq. (\ref{linear7}) into Eq. (\ref{linear8}), we can
obtain the third transformation (from K to K$''$)
\begin{equation}
\left\{
\begin{array}{ll}
&  x''=\frac{1-\lambda_{1} \omega}{\sqrt{1+\lambda_{2} \omega}\sqrt{1+\lambda_{1} v}}\left(x-\frac{v+\omega}{1-\lambda_{1} \omega}t\right),                                 \\
&   t''=\frac{1-\lambda_{2} v}{\sqrt{1+\lambda_{2}
\omega}\sqrt{1+\lambda_{1}
v}}\left(t+\frac{\lambda_{1}+\lambda_{2}}{1-\lambda_{2}v}x\right).
\end{array}
\right. \label{linear9}
\end{equation}
It follows that Eq. (\ref{linear9}) should be rewritten in the
form
\begin{equation}
\left\{
\begin{array}{ll}
&     x''=\frac{1}{\sqrt{1+\Lambda V}}(x-Vt),                 \\
&    t''=\frac{1}{\sqrt{1+\Lambda V}}(t+\Lambda x),
\end{array}
\right. \label{linear10}
\end{equation}
where $V$ denotes the relative velocity between frames K and
K$''$. Comparing Eq. (\ref{linear10}) with Eq. (\ref{linear9}), we
can obtain the expressions for parameters $V$ and $\Lambda$:
\begin{equation}
V=\frac{v+\omega}{1-\lambda_{1} \omega},  \quad
\Lambda=\frac{\lambda_{1}+\lambda_{2}}{1-\lambda_{2}v}.
\label{LamV}
\end{equation}
Moreover, the following relation
\begin{equation}
\frac{1-\lambda_{1} \omega}{\sqrt{1+\lambda_{2}
\omega}\sqrt{1+\lambda_{1} v}}=\frac{1-\lambda_{2}
v}{\sqrt{1+\lambda_{2} \omega}\sqrt{1+\lambda_{1} v}}
\label{equal}
\end{equation}
should also be fulfilled. This means that $1-\lambda_{1}
\omega=1-\lambda_{2} v$, or
\begin{equation}
\frac{\lambda_{1}}{v}=\frac{\lambda_{2}}{\omega}.    \label{ratio}
\end{equation}
Since the terms on the left- and right-handed sides in
(\ref{ratio}) are the parameters corresponding to their respective
transformations (\ref{linear7}) and (\ref{linear8}), they should
be independent of each other. Keep in mind that the two terms in
(\ref{ratio}) are equal to each other. Thus they should be equal
to a constant number, say, $q$, {\it i.e.},
\begin{equation}
\frac{\lambda_{1}}{v}=\frac{\lambda_{2}}{\omega}=q,
\label{ratio2}
\end{equation}
and subsequently
\begin{equation}
\lambda_{1}=qv,   \quad    \lambda_{2}=q\omega.     \label{ratio3}
\end{equation}
Insertion of the above expressions into Eq. (\ref{LamV}) yields
\begin{equation}
V=\frac{v+\omega}{1-qv\omega},    \quad
\Lambda=\frac{q(v+\omega)}{1-q\omega v},    \label{LamV2}
\end{equation}
where the first formula is the law for the addition of velocities,
and the second formula can be rewritten as
\begin{equation}
\Lambda=qV.      \label{qV}
\end{equation}
Such a form is consistent with expression (\ref{ratio3}). Further
calculation can show that both of the two terms on the left- and
right-handed sides of Eq. (\ref{equal}) are equal to the
transformation coefficient ${1}/{\sqrt{1+\Lambda V}}$ in
transformation (\ref{linear10}). This, therefore, means that the
above formulation is self-consistent.

Thus, according to Eqs. (\ref{lambda}) and (\ref{linear4}), the
ultimate form of the spacetime coordinate transformation from K to
K$'$ is given by
\begin{equation}
\left\{
\begin{array}{ll}
&  x'=\frac{1}{\sqrt{1+qv^{2}}}(x-vt),                                 \\
&   t'=\frac{1}{\sqrt{1+qv^{2}}}(t+qvx),
\end{array}
\right. \label{gongtong}
\end{equation}
and the law for the addition of velocities is
\begin{equation}
u'=\frac{u-v}{1+qvu}.     \label{add}
\end{equation}

It is apparently seen that if the parameter $q=0$, then Eqs.
(\ref{gongtong}) and (\ref{add}) are both reduced to the forms in
Newtonian mechanics. In other words, the Galilean transformation
is the most simple coordinate form that fulfills the principle of
relativity. However, the case of $q\neq 0$ also satisfies the
principle of relativity and is permitted to exist in physics. In
the subsection that follows, we discuss the physical meanings of
the parameter $q$, and then based on Eq. (\ref{gongtong}), show
that the Galilean and Lorentz transformations are the only two
self-consistent coordinate transformations.

\subsection{The existence of an invariant velocity}
It is clear that transformation (\ref{gongtong}) has a form
analogous to the Lorentz transformation, if the parameter $q\neq
0$. One of the logical conclusions in the above formulation is
that there exists an invariant velocity in formula (\ref{add}) for
the addition of velocities: specifically, if the invariant
velocity is $\xi$ (such an invariant velocity is a same constant
in both frames K and K$'$, {\it i.e.}, $u=\xi$, $u'=\xi$), then
from Eq. (\ref{add}) we can have
\begin{equation}
\xi=\frac{\xi-v}{1+qv\xi},     \label{add2}
\end{equation}
and the following relation
\begin{equation}
q\xi^{2}=-1
\end{equation}
can be derived. Thus the invariant velocity can be expressed in
terms of the constant $q$, {\it i.e.},
\begin{equation}
\xi=\sqrt{-\frac{1}{q}}.
\end{equation}
By using the relation
\begin{equation}
q=-\frac{1}{\xi^{2}},     \label{qq}
\end{equation}
the coordinate transformation (\ref{gongtong}) and the law
(\ref{add}) for velocity addition can be rewritten as
\begin{equation}
\left\{
\begin{array}{ll}
&  x'=\frac{1}{\sqrt{1-\frac{v^{2}}{\xi^{2}}}}\left(x-vt\right),                                 \\
&
t'=\frac{1}{\sqrt{1-\frac{v^{2}}{\xi^{2}}}}\left(t-\frac{vx}{\xi^{2}}\right),
\end{array}
\right. \label{gongtong2}
\end{equation}
and
\begin{equation}
u'=\frac{u-v}{1-\frac{vu}{\xi^{2}}},     \label{add22}
\end{equation}
where the invariant velocity $\xi$ is involved.

Note that Eq. (\ref{gongtong2}) has a same form as the Lorentz
transformation. Then what about the numerical value of the
invariant velocity $\xi$? It should be determined by the
experiments. Modern experiments show that the invariant velocity
$\xi$ has the same value as the speed of light in vacuum.

In the above we derive the Lorentz transformation without the
postulate of constancy of the speed of light. In contrast, such a
postulate can be viewed as a necessary consequence of the
derivation. Both the Galilean and Lorentz transformations, which
correspond to the different invariant velocities, can be derived
in the above formulation.

\section{The Edwards transformation and the principle of permutation invariance}
In the preceding section, we adopted Einstein's definition of
simultaneity (standard synchrony) and assumed that the one-way
velocity is an observable quantity. However, as stated in
Introduction, Einstein's standard synchrony is simply one of the
various possible definitions of simultaneity. In this section, we
make use of the principle of permutation invariance and establish
the generalized coordinate transformations, which correspond to
other definitions of relativity of simultaneity and agree with the
postulate of relativity as well as the principle of causality.

\subsection{Transformation coefficients independent of the velocities of the test particle}

The principle of permutation invariance presented in this paper is
equivalent to the postulate of relativity: specifically, the
permutation operation can guarantee that the physical phenomena in
all inertial frames of reference occur in an identical manner. In
what follows, we show how the principle of permutation invariance
works in deriving the Edwards transformation. Assume that frame
K$'$ moves relative to K in the positive $\hat{x}$-direction with
velocity $v$, while K moves relative to K$'$ with velocity $-v'$.
The space coordinate transformation between frames K and K$'$ is
\begin{equation}
\left\{
\begin{array}{ll}
&  x'=k(x-vt),                                 \\
&   x=k'(x'+v't').
\end{array}
\right. \label{linearr}
\end{equation}
Substitution of the equation of motion ({\it i.e.}, $x=ut$,
$x'=u't'$) of a test particle into Eq. (\ref{linearr}) yields
\begin{equation}
k'k=\frac{u'u}{(u-v)(u'+v')}.       \label{kk}
\end{equation}
As $k'k$ should be independent of the particle velocity $u$, we
have ${\rm d}(k'k)/{\rm d}u=0$, and then obtain
\begin{equation}
\frac{{\rm d}u'}{{\rm d}u}=\frac{(u'+v')u'v}{(u-v)uv'}.
\end{equation}
The solution to this equation is
\begin{equation}
u'=\frac{u-v}{\frac{v}{v'}+\lambda u},      \label{uprimed}
\end{equation}
where $\lambda$ is a parameter independent of the particle
velocity $u$. Substitution of expression (\ref{uprimed}) into Eq.
(\ref{kk}) yields
\begin{equation}
k'k=\frac{1}{1+\lambda v'}.    \label{kk2}
\end{equation}
Alternatively, by using the permutation operation
($k'\leftrightarrow k$, $v'\leftrightarrow -v$,
$\lambda\leftrightarrow \lambda'$) and the principle of
permutation invariance, we can express $k'k$ in terms of $v$, {\it
i.e.},
\begin{equation}
k'k=\frac{1}{1+\lambda' (-v)}.   \label{kk2Alter}
\end{equation}
In the permutation operation, the symbol $\leftrightarrow$ means
the interchange of two quantities on its two sides. By comparing
expression (\ref{kk2Alter}) with (\ref{kk2}), one can see that if
$\lambda=\varsigma v$, $\lambda'=-\varsigma v'$, then we have the
relations $\lambda v'=-\lambda'v'=\varsigma v'v$. Here the
$\varsigma$ is a parameter that is invariant under the permutation
transformation. Thus, both expressions (\ref{kk2Alter}) and
(\ref{kk2}) can be rewritten as
\begin{equation}
k'k=\frac{1}{1+\varsigma v'v},       \label{determinedkk}
\end{equation}
which is invariant under the permutation operation.

According to Eq. (\ref{linearr}), the spacetime coordinate
transformation from K to K$'$ is given by
\begin{equation}
\left\{
\begin{array}{ll}
&  x'=k(x-vt),                                 \\
&   t'=k\left(\frac{v}{v'}t+\varsigma vx\right).
\end{array}
\right. \label{linear}
\end{equation}
Under such a transformation, the law for addition velocities is
\begin{equation}
u'=\frac{u-v}{\frac{v}{v'}+\varsigma uv}.     \label{additionlaw}
\end{equation}

By using the principle of permutation invariance
($k\leftrightarrow k'$, $v\leftrightarrow -v'$, $x\leftrightarrow
x'$, $t\leftrightarrow t'$), we can obtain the inverse
transformation
\begin{equation}
\left\{
\begin{array}{ll}
&  x=k'(x'+v't'),                                 \\
&   t=k'\left(\frac{v'}{v}t'-\varsigma v'x'\right).
\end{array}
\right. \label{linearrr}
\end{equation}
In the meanwhile, from Eq. (\ref{additionlaw}), we can obtain the
inverse transformation for the addition of velocities
\begin{equation}
u=\frac{u'+v'}{\frac{v'}{v}-\varsigma u'v'}
\end{equation}
by using the permutation operation ($v\leftrightarrow -v'$,
$u\leftrightarrow u'$).

\subsection{Determination of parameters by permutation operation}

In transformations (\ref{linear}) and (\ref{linearrr}) the only
retained parameters (and functions), which should be determined,
are $v/v'$, $k, k'$ and $\varsigma$. In the following discussions,
we can obtain these parameters (and functions) by using the
principle of permutation invariance (covariance):

i) it is found that the functions $v/v'$ and $v'/v$ should take
the following forms
\begin{equation}
\frac{v}{v'}=\frac{1}{1-\eta v'},    \qquad
\frac{v'}{v}=\frac{1}{1+\eta v},       \label{vdividev}
\end{equation}
where $\eta$ is a permutation-invariance parameter that will be
determined below. These two expressions satisfy the principle of
permutation covariance: by using the permutation $v\leftrightarrow
-v'$, $v'\leftrightarrow -v$, the second expression can be
rewritten as the first one, and vice versa. This means that the
two expressions in (\ref{vdividev}) are the only self-consistent
choices for the functions $v/v'$ and $v'/v$.

ii) it follows from Eq. (\ref{determinedkk}) that the
transformation coefficients $k$ and $k'$, which agrees with the
principle of permutation covariance, should have the following
forms
\begin{equation}
\left\{
\begin{array}{ll}
&      k'=\sqrt{\left(\frac{v}{v'}\right)^{\sigma}\frac{1}{1+\varsigma v'v}},                             \\
&
       k=\sqrt{\left(\frac{v'}{v}\right)^{\sigma}\frac{1}{1+\varsigma
v'v}}.
\end{array}
\right. \label{linearrrr}
\end{equation}
Clearly, the second expression can be transformed into the first
one, and vice versa under the permutation transformation
($k\leftrightarrow k'$, $v\leftrightarrow -v'$).

iii) there are various choices for the form of the parameter
$\eta$ that is invariant under the permutation operation. As one
of the most simple forms, $\eta$ can be chosen as
\begin{equation}
\eta=\frac{X+X'}{c}.
\end{equation}
The physical meanings of the parameters $X, X'$ and $c$ will be
revealed in what follows.

Let us consider a round-trip motion of a test particle in K and
K$'$. If the particle is moving from point A to point B in frame K
with a velocity $c_{+}$ parallel to the positive
$\hat{x}$-direction, and then by reflection, back to point A with
a velocity $-c_{-}$ parallel to the negative $\hat{x}$-direction.
In the meanwhile, the same particle is moving from point A$'$ to
point B$'$ in frame K$'$ with a velocity $c'_{+}$ parallel to the
positive $\hat{x'}$-direction, and by reflection, back from point
B$'$ to point A$'$ with $-c'_{-}$ parallel to the negative
$\hat{x'}$-direction. According to expression (\ref{additionlaw}),
one can obtain the relations between the to-and-fro velocities
$c_{+}, c_{-}$ and $c'_{+}, c'_{-}$, {\it i.e.},
\begin{eqnarray}
 c'_{+}&=&\frac{c_{+}-v}{\frac{v}{v'}+\varsigma c_{+}v},            \nonumber \\
 -c'_{-}&=&\frac{(-c_{-})-v}{\frac{v}{v'}+\varsigma
(-c_{-})v}.        \label{c+c-}
\end{eqnarray}
Then by using the principle of permutation invariance, one can
obtain a relation
\begin{eqnarray}
& &   v\left[\left(\frac{1}{c'_{+}c_{+}}-\frac{1}{c'_{-}c_{-}}\right)+\frac{X+X'}{c}\left(\frac{1}{c_{+}}+\frac{1}{c_{-}}\right)\right]                \nonumber \\
&=&
v'\left[\left(\frac{1}{c'_{+}c_{+}}-\frac{1}{c'_{-}c_{-}}\right)+\frac{X+X'}{c}\left(\frac{1}{c'_{+}}+\frac{1}{c'_{-}}\right)\right].
\label{Vrelation}
\end{eqnarray}
It follows that the relations
\begin{equation}
\left\{
\begin{array}{ll}
&    c'_{+}=\frac{c}{1-X'}                          \\
&     c'_{-}=\frac{c}{1+X'}
\end{array}
\right.
 \qquad \left\{
\begin{array}{ll}
&    c_{+}=\frac{c}{1-X}                    \\
&     c_{-}=\frac{c}{1+X}
\end{array}
\right.
 \label{SolutionOfc+}
\end{equation}
satisfy Eq. (\ref{Vrelation}). It is shown from expression
(\ref{SolutionOfc+}) that the velocities $c_{+}, c_{-}$ and
$c'_{+}, c'_{-}$ agree with the following relation
\begin{equation}
\frac{1}{2}\left(\frac{1}{c'_{+}}+\frac{1}{c'_{-}}\right)=\frac{1}{2}\left(\frac{1}{c_{+}}+\frac{1}{c_{-}}\right)=\frac{1}{c}.
\label{twowayspeed}
\end{equation}
It can be verified that for Eq. (\ref{Vrelation}), there exist
many solutions other than (\ref{SolutionOfc+}) for $c_{+}, c_{-}$
and $c'_{+}, c'_{-}$, and that the solution (\ref{SolutionOfc+})
that is one of the most simplest ones corresponds to the massless
particles and the other more complicated solutions belong to the
massive particles. If such a massless particle is photon, then the
physical meaning of relation (\ref{twowayspeed}) is the constancy
of the two-way speed of a light: specifically, the average speed
of a light pulse travelling from point A to point B and by
reflection, back to A is an invariant quantity that is independent
of the choice of the inertial reference frames. Here $c$ denotes
the two-way average speed of light in free vacuum. Clearly, since
$X$ and $X'$ are adjustable parameters relying upon the means of
clock synchronization, there is no absolute simultaneity
relations, and the standard Einsteinian synchrony is simply the
choice corresponding to the postulate that no causal influence can
travel faster than the speed of light in vacuum.

With the help of Eqs. (\ref{c+c-}) and (\ref{SolutionOfc+}) we can
obtain the explicit expression for the permutation-invariance
parameter $\varsigma$
\begin{equation}
\varsigma=\frac{1}{c^{2}}\left(X^{2}-1\right)+\frac{X-X'}{cv}.
\label{varsigma}
\end{equation}
By using the permutation operation ($X\leftrightarrow X'$,
$v\leftrightarrow -v'$), we can obtain an alternative expression
\begin{equation}
\varsigma=\frac{1}{c^{2}}\left(X'^{2}-1\right)+\frac{X-X'}{cv'}.
\label{varsigma2}
\end{equation}
It can be readily verified that the parameter $\varsigma$ in
expression (\ref{varsigma2}) is truly equal to that in expression
(\ref{varsigma}). This means that the principle of permutation
invariance presented in this paper is self-consistent. Thus, from
Eqs. (\ref{linearrrr}), (\ref{varsigma}) and (\ref{varsigma2}),
the explicit expression for $k'k$ is given by
\begin{eqnarray}
k'k&=&\frac{1}{1+\left[\frac{1}{c^{2}}\left(X^{2}-1\right)+\frac{X-X'}{cv}\right]v'v}       \nonumber \\
&=&
\frac{1}{1+\left[\frac{1}{c^{2}}\left(X'^{2}-1\right)+\frac{X-X'}{cv'}\right]v'v}.
\end{eqnarray}

The coordinate transformation (\ref{linear}) and its inverse
transformation (\ref{linearrr}) with the coefficients $k, k'$
defined as (\ref{linearrrr}) can be viewed as the generalized
Edwards transformations. The Edwards transformation suggested in
1963 is in fact a simple one, the parameter $\sigma$ of which is
$\sigma=1$.
\subsection{The Edwards transformation}
If the parameter $\sigma$ is taken to be $1$, then the
transformation coefficients $k, k'$ in Eqs. (\ref{linear}) and
(\ref{linearrr}) are of the form
\begin{equation}
k=\frac{1}{\sqrt{\left(1+\frac{v}{c}X\right)^{2}-\frac{v^{2}}{c^{2}}}},
\quad
k'=\frac{1}{\sqrt{\left(1-\frac{v'}{c}X'\right)^{2}-\frac{v'^{2}}{c^{2}}}}.
\end{equation}
Thus the coordinate transformation with the constancy of the
two-way average speed of light reads
\begin{equation}
\left\{
\begin{array}{ll}
&     x'=k(x-vt)                              \\
&
t'=k\left[\left(1+\frac{X+X'}{c}v\right)t+\left(\frac{X^{2}-1}{c^{2}}+\frac{X-X'}{cv}\right)vx\right].
\end{array}
\right. \label{Edwards1}
\end{equation}
The corresponding inverse transformation is
\begin{equation}
\left\{
\begin{array}{ll}
&     x=k'(x'+v't')                              \\
&
t=k'\left[\left(1-\frac{X+X'}{c}v'\right)t'-\left(\frac{X'^{2}-1}{c^{2}}+\frac{X-X'}{cv'}\right)v'x'\right].
\end{array}
\right. \label{Edwards2}
\end{equation}
Apparently, the inverse transformation (\ref{Edwards2}) can be
obtained from (\ref{Edwards1}) by using the permutation operation.

\section{Connection between Edwards transformation and general coordinate transformation}
The Edwards transformation (\ref{Edwards1}) can be reduced to the
Lorentz transformation if the Edwards parameters $X=X'=0$. In the
Edwards transformation, Einstein's postulate of constancy of the
one-way speed of light is replaced by the principle of constancy
of the two-way average speed of light. In what follows we point
out the connection between the Edwards transformation and the
general coordinate transformation. The to-and-fro speeds of light
in the Edwards spacetime are $c_{+}=c/(1-X)$ and
$-c_{-}=-c/(1+X)$, respectively. It is easily seen that the two
speeds, $c_{+}$ and $-c_{-}$, of the light fulfill the following
quadratic equation
\begin{equation}
\left(1-X^{2}\right)u^{2}-2Xcu-c^{2}=0,
\label{quadraticequation}
\end{equation}
where $u$ is defined by $u={\rm d}x/{\rm d}t$. Apparently, the
solutions of Eq. (\ref{quadraticequation}) are $u_{1}=c_{+}$,
$u_{2}=-c_{-}$. Eq. (\ref{quadraticequation}) can be rewritten as
\begin{equation}
\left(1-X^{2}\right){\rm d}x^{2}-2X{\rm d}xc{\rm d}t-c^{2}{\rm
d}t^{2}=0,     \label{quadraticequation2}
\end{equation}
and the matrix form for Eq. (\ref{quadraticequation2}) reads
\begin{equation}
 \left({\rm d}x^{0} \quad {\rm d}x\right)\left( {\begin{array}{*{20}c}
   {-1} & {-X}  \\
   {-X} & {1-X^{2}}  \\
\end{array}} \right)\left( {\begin{array}{*{20}c}
   {{\rm d}x^{0}}  \\
   {{\rm d}x}  \\
\end{array}} \right)=0,
\label{metriccc}
\end{equation}
where ${\rm d}x^{0}=c{\rm d}t$. It can be further rewritten as a
square of spacetime interval, {\it i.e.},
\begin{equation}
{\rm d}s^{2}=g_{\mu\nu}{\rm d}x^{\mu}{\rm d}x^{\nu}=0,
\label{interval}
\end{equation}
where $g_{\mu\nu}$ is a spacetime metric tensor,
\begin{equation}
g_{\mu\nu}=\left( {\begin{array}{*{20}c}
   {-1} & {-X}  \\
   {-X} & {1-X^{2}}  \\
\end{array}} \right).
\label{metric}
\end{equation}
Although the tensor $g_{\mu\nu}$ seems to be a metric of curved
spacetime, it is actually a flat metric since the Edwards
parameter $X$ is constant for a certain reference frame, and
therefore all the components of the Riemannian curvature tensor
vanish, {\it i.e.}, $R_{\mu\nu\alpha\beta}=0$. This means that the
spacetime with such a line element (\ref{interval}) is a flat
spacetime rather than a curved one, and that the symmetric tensor
(\ref{metric}) can be changed into a diagonal flat metric tensor
by using a general coordinate transformation. It is thus believed
that the Edwards transformation is different from the Lorentz
transformation only by a certain general coordinate
transformation. Note that such a general coordinate transformation
is a global (rather than local) transformation. Therefore, the
Edwards transformation is equivalent to the Lorentz
transformation. The special relativity with the constancy of the
two-way average speed of light predicts the same observable
effects as Einstein's special relativity did.

For the Edwards spacetime, the physical meaning of the Edwards
parameter $X$ in expression (\ref{metric}) is analogous to a
gravitational (gravitomagnetic) potential \cite{Shen}. However, as
such a gravitomagnetic potential is a constant number, it does not
lead to any force field effects for any observable physical
quantities. For example, in both the Lorentz and Edwards
transformations, the two-way average speed of light, which is an
observable quantity, takes the same value. But the values for the
one-way speed of light, which is not an observable quantity, are
different in the Lorentz and Edwards transformations. Such a
difference is due to the so-called synchronization gauge ({\it
i.e.}, the arbitrariness in the definition of relativity of
simultaneity).
\section{Concluding remarks}

As we have no ideal means of clock synchronization ({\it e.g.},
the infinite-speed signals), there are no absolute and unique
definitions of simultaneity. Einstein's simultaneity (standard
synchrony) is simply a special case among various possible
definitions of simultaneity, which obey the principle of
causality. The Lorentz transformation corresponds to the one of
the most simplest means of clock synchronization, and the
relativity of simultaneity in Einstein's special relativity
depends upon such a special choice of synchronization. The
measurement of the quantities, which cannot be directly
observable, has a close relation to the definition of simultaneity
(synchronization gauge). Different definitions of simultaneity
will give different results of measurement for, {\it e.g.}, the
one-way speed of light. But for any observable quantities,
different definitions of simultaneity (and various clock
synchronizations) will lead to the same measurement results.

The Edwards transformation was derived based on the principle of
permutation invariance (covariance), which incarnates both the
principle of relativity and the arbitrariness in definition of
simultaneity. As there is an effective gravitomagnetic potential
(the Edward parameter), the Edward spacetime can be considered a
Riemannian spacetime rather than a Minkowski spacetime. But such a
gravitomagnetic potential in the Edward spacetime is constant
(independent of the spacetime coordinates), so that there exists a
global coordinate transformation, which can transform the Edward
spacetime into the Minkowski spacetime. In this sense, the Edward
spacetime is equivalent to the Minkowski spacetime.

\begin{acknowledgments}
This work is supported in part by the Zhejiang Provincial Natural
Science Foundation (China) under Project No. Y404355.
\end{acknowledgments}

\end{document}